# The Proof-Theoretic Foundations of the Principle of Reiteration in Natural Deduction

**Khashayar Irani**
**Birkbeck College, University of London**
**k.irani@mathematicallogic.com**
**October 2026**

**Abstract**

This paper investigates the proof-theoretic foundations of the principle of reiteration (R) in natural deduction, arguing that its legitimacy originates in the structural preservation of formulae within sequent calculus rather than in an independent logical rule.[1] By examining singlesuccedent and multisuccedent sequent calculus, we demonstrate how logical operations transform principal formulae while preserving unaffected antecedent and succedent contexts. We subsequently analyse the schematic formulation of reiteration and compare natural deduction derivations with corresponding sequent calculus derivations. This comparison identifies two fundamental functions of reiteration, namely the reuse of previously employed formulae and the restoration of previously derived conclusions. We conclude that reiteration introduces no independent inferential content but explicitly represents the structural continuity of inferential dependencies already included into sequent calculus.

## 1. Introduction

The principle of reiteration (R) occupies a distinctive position within natural deduction, particularly concerning the continued availability of formulae that have previously participated in inferential operations.[2] Unlike the introduction and elimination rules

---

[1] This enquiry, together with a number of other investigations that will be published either prior to or alongside it, proposes several structural analyses concerning sequent calculus and natural deduction that we believe have been overlooked in both earlier and contemporary works in proof theory. The central thesis of the present paper was originally inspired by certain proof-theoretic analyses developed in Khashayar Irani (2025a), which prompted us to examine the principle of reiteration in natural deduction derivations. In the future, we aim to produce a comprehensive manuscript that brings together the results of this work and a few other similar proof-theoretic studies and presents them as fundamental structural principles of sequent calculus and natural deduction systems.

[2] In this study, our formulation and presentation of natural deduction follow the Gentzen-Prawitz style, with the exception of employing the principle of reiteration, which we have not encountered in the works of these authors. For a general understanding of the Gentzen-Prawitz tradition, see Gerhard Gentzen, as cited in M. E. Szabo (1969), and Dag Prawitz (1965 & 1971). Furthermore, as Francis Jeffry Pelletier & Allen Hazen (2026) explain, natural deduction is not confined to the Gentzen-Prawitz tradition but encompasses alternative

governing logical constants, reiteration does not introduce new logical content but permits an already available formula to be restated at a subsequent stage of a derivation. Its proof-theoretic legitimacy yet requires examination, especially when a formula that has participated in an inference is subsequently employed in another. The central question is whether this continued availability constitutes an independent feature of natural deduction or can be explained through the structural behaviour of sequent calculus.[3] This manuscript defends the latter position, arguing that reiteration derives its legitimacy from the preservation of formulae within sequent calculus derivations. In particular, it distinguishes between the inferential transformation of principal formulae and the structural preservation of those formulae that remain unaffected by a given logical operation. Within sequent calculus, this distinction is represented through the antecedent-succedent structure, whereas natural deduction operates within a comparatively nonlocal derivational atmosphere in which previously employed assumptions and derived conclusions may remain available for subsequent inferential operations.[4] To examine this correspondence, we consider singlesuccedent and multisuccedent sequent calculus derivations before comparing particular applications of the rules →E and ∧I:

$$\frac{\mathbf{A \to B} \qquad \mathbf{[A]}}{\mathbf{B}}\,\mathbf{{\to}E} \qquad\qquad \frac{\mathbf{A} \qquad \mathbf{B}}{\mathbf{A \land B}}\,\mathbf{{\land}I}$$

as well as reiteration in natural deduction with their corresponding PRINCIPLES NAMELY, L→ and L∧:

$$\frac{\mathbf{\Gamma \Rightarrow A} \qquad \mathbf{B.\Gamma \Rightarrow C}}{\mathbf{A \to B.\Gamma \Rightarrow C}}\,\mathbf{L{\to}} \qquad\qquad \frac{\mathbf{A,B,\Gamma \Rightarrow C}}{\mathbf{A \land B,\Gamma \Rightarrow C}}\,\mathbf{L{\land}}$$

in sequent calculus.[5] This comparison serves to explain both the reuse of previously employed formulae and the restoration of previously derived conclusions. Two potential objections are also examined, concerning the legitimacy of reusing formulae after their earlier inferential employment and the correspondence between the different conjunction rules in the

---

formulations, including Stanisław Jaśkowski's system of subordinate proofs, Frederic Brenton Fitch's system and the Suppes–Lemmon style. These approaches differ in their presentation of derivations, treatment of assumptions and organisation of inferential rules. For a comprehensive examination of these systems and their formal differences, see Pelletier & Hazen (2026).

[3] For a detailed study of sequent calculus, particularly its logical, structural and other proof-theoretic aspects, see Anne Sjerp Troelstra & Helmut Schwichtenberg (2000), Sara Negri & Jan von Plato (2001 & 2011), and Irani (2025b).

[4] For an analysis of the nonlocal nature of natural deduction and the freedom to introduce assumptions at any previous inferential stage, see Irani (2026a).

[5] For a proof-theoretic and systematic understanding of multiconclusion natural deduction and multisuccedent sequent calculus systems, respectively, see Timothy Smiley & D. J. Shoesmith (1978), Troelstra & Schwichtenberg (2000), and Negri & von Plato (2001 & 2011). Additionally, for a recent formulation of a classical multisuccedent sequent calculus framework named GM-Calculus, see Irani (2025b).

two proof environments. The analysis thus aims to demonstrate that reiteration is not a primitive logical rule but an explicit expression of structural preservation already implicit in sequent calculus. However, before developing this account further, it is important to consider a number of earlier and contemporary investigations that have examined reiteration directly or explored related structural properties of natural deduction and sequent calculus.

The proof-theoretic status of reiteration has received attention in earlier treatments of natural deduction, whilst contemporary investigations of structural rules and inferential dependencies provide further perspectives relevant to its analysis. An important early discussion appears in Fitch (1952), who characterises reiteration as the repetition of a proposition from an enclosing proof within a subordinate proof. Significantly, Fitch distinguishes reiteration from the introduction and elimination rules, maintaining that it is not properly a rule of direct consequence, since its operation concerns the occurrence of a proposition in distinct proofs rather than an inferential relation between propositions within the same proof. This distinction raises the question of whether reiteration should be understood as an independent logical operation or as a feature of the structural organisation of derivations. More recently, Yazeed Alrubyli (2021), in presenting a natural deduction calculus for first order logic, explicitly includes reiteration as a means of reproducing an earlier step within a proof or introducing it into a subordinate proof. Although this treatment identifies its practical function, it does not investigate the sequent calculus structures that might justify its employment. A different but related perspective emerges in Greg Restall (2023), who examines structural rules in natural deduction with alternatives and their correspondence with sequent calculus. By distinguishing the structural organisation of assumptions from the introduction and elimination rules governing logical connectives, Restall demonstrates how variations in assumption discharge, including duplicate and vacuous discharge, correspond to differences in structural principles. His treatment of assumption classes, together with the structural operations of storing and retrieving alternatives, illustrates how the availability and employment of formulae depend upon the organisation of proofs. Similarly, Junyoung Jang, Sophia Roshal, Frank Pfenning and Brigitte Pientka (2024) develop a natural deduction formulation of adjoint logic corresponding to its sequent calculus presentation, examining how weakening, contraction and restrictions upon assumption usage are governed by the structural properties of different modes. Their treatment of contexts demonstrates that the repeated employment of assumptions is not universally permissible but depends upon the structural conditions of the underlying calculus. Neither Restall nor Jang, Roshal, Pfenning and Pientka directly establishes a general justification of reiteration. But these investigations collectively indicate the importance of distinguishing logical inference from the structural conditions governing the continued availability of formulae. The present study develops this distinction specifically for reiteration, arguing that its two functions, the reuse of previously employed formulae and the restoration of previously derived conclusions, can be explained through the preservation of antecedent and succedent contexts in sequent calculus derivations.

The significance of this research extends to the proof-theoretic treatment of double negation introduction (DNI: $A \vdash \neg\neg A$) and double negation elimination (DNE: $\neg\neg A \vdash A$), originally presented in Irani (2025a), particularly where the preservation of previously derived conclusions is required.[6] Although DNI can be represented within a singleconclusion natural deduction framework:

$$\cfrac{\cfrac{[A] \qquad [\neg A]}{\bot}\,\neg E}{\neg\neg A}\,\neg I$$

without employing reiteration, the proposed derivation of DNE:

$$\cfrac{\cfrac{\cfrac{[A]}{A \lor B[}\,\lor I \qquad [(A \lor B) \to A]}{A}\,\to E \qquad \cfrac{\cfrac{\cfrac{\cfrac{[A]}{A \lor B}\,\lor I \qquad [\neg(A \lor B)]}{\bot}\,\neg E}{\neg A}\,\neg I \qquad [\neg\neg A]}{\bot}\,\neg E}{A}\,R$$

considered in the underlying analysis possesses a multiconclusion structure in which a previously and independently derived formula A must remain available throughout a reductio operation before being restated as the final conclusion. The legitimacy of this application therefore depends upon explaining why a formula remains available notwithstanding the subsequent performance of other inferential operations. However, the present examination does not undertake a separate derivation of either DNI or DNE, since its immediate objective is to establish the proof-theoretic foundations of reiteration independently of these particular applications. This distinction is methodologically important because the structural function of reiteration must be distinguished from the logical operations in which previously derived formulae participate. The fact that reiteration introduces no additional inferential content does not diminish its proof-theoretic significance. Rather, its importance lies in making explicit the continued availability of formulae whose preservation is already represented within the structure of sequent calculus. As a result, the employment of reiteration in a derivation involving reductio does not itself perform the reductio operation or introduce an additional logical conclusion. Instead, it permits a formula derived independently of that operation to be restated when required. This understanding also clarifies why reiteration should not be regarded as concealing the complexities of classical derivation, since its function is confined to preserving and restating what has already been derived. The distinction between structural preservation and inferential transformation thus provides the conceptual basis for examining the correspondence between natural deduction and sequent calculus. By establishing this distinction before considering particular derivations, the analysis seeks to explain the proof-theoretic legitimacy of reiteration without attributing to it an independent inferential capacity or introducing logical principles beyond those already represented within the respective proof environments.

[6] In Irani (2025a), we provided a comprehensive investigation into the claim that the proof-theoretic origins of DNI and DNE are rooted in reductio ad absurdum (RAA) derivations. Equally, a supplementary study concerning a general theory of RAA can be found in Irani (2026b).

## 2. The Proof-Theoretic Foundations of Reiteration

After distinguishing the structural function of reiteration from the logical operations in which previously derived formulae participate, we may now examine its proof-theoretic foundations through the formal organisation of sequent calculus derivations. In sequent calculus, logical rules operate upon designated principal formulae, while the remaining formulae constitute the inferential context within which those operations are performed. This distinction is particularly significant because the application of a logical rule does not entail the transformation or disappearance of every formula occurring within the sequent. Instead, formulae that do not participate in the principal operation are preserved according to the contextual structure of the rule. Thus, when an inference operates upon an antecedent formula, the succedent remains unchanged unless the rule explicitly requires otherwise. Similarly, within multisuccedent sequent calculus, an operation directed towards a particular succedent formula may leave the remaining succedent formulae unaffected. Such preservation is not an additional logical inference but a structural feature of the rule itself, whereby the surrounding context is maintained throughout the transition from premises to conclusion. This distinction provides a formal explanation of how inferential operations can proceed without requiring the independent rederivation of formulae that remain available within the proof. While natural deduction and sequent calculus represent inferential dependencies differently, the structural preservation of formulae in the latter offers a basis for understanding the continued availability of assumptions and previously derived conclusions in the former. In particular, reiteration explicitly permits the restatement of an available formula, whereas the corresponding sequent calculus representation may preserve that formula within the antecedent or succedent without applying a separate rule for its reproduction. To examine this relationship systematically, we shall initially consider a singlesuccedent derivation in which an antecedent formula undergoes a logical transformation while the succedent remains unchanged. We shall subsequently extend the analysis to multisuccedent derivations, where the preservation of formulae can also be demonstrated when a logical operation concerns the succedent itself. These cases will provide the formal groundwork for comparing the structural behaviour of sequent calculus with the two functions of reiteration identified in the abstract.

Consider the following singlesuccedent sequent calculus derivation, in which the rule L∧:

$$\frac{A, B, \Gamma \Rightarrow C}{A \wedge B, \Gamma \Rightarrow C}\, L\wedge$$

operates upon the antecedent formula A ∧ B while preserving the succedent formula C. In this inference, the constituent formulae A and B occur separately in the antecedent of the premise and are combined into the principal formula A ∧ B in the conclusion. The operation therefore concerns the logical structure of conjunction, whereas the remaining antecedent context Γ and the succedent formula C are preserved without alteration. This distinction is important because the transformation of A and B does not affect the occurrence of C, which continues to occupy the succedent position after the application of L∧. Consequently, the rule requires neither a separate derivation of C nor an additional inferential operation to secure its continued occurrence. The preservation of C is included into the formal structure of the inference itself, independently of the logical transformation performed upon the antecedent. Furthermore, the example illustrates that the preservation of a conclusion does not depend upon that conclusion participating as a principal formula in the subsequent logical operation. Rather, C remains available precisely because the application of L∧ concerns the antecedent conjunction and leaves the succedent unchanged. This feature provides an initial basis for understanding the proof-theoretic function of reiteration in natural deduction. When a previously derived conclusion must be restated after other inferential operations have taken place, reiteration makes its continued availability explicit without introducing a new logical transformation. In the corresponding sequent calculus representation, however, the preservation of the succedent formula is already expressed through its unchanged occurrence across the inference. The relevant correspondence therefore concerns the structural continuity of the conclusion rather than an identity between the logical rules employed in the two proof surroundings. Although this singlesuccedent example demonstrates preservation when the principal formula occurs in the antecedent, it does not yet establish how the same phenomenon operates when a rule transforms a succedent formula. That further question requires consideration of multisuccedent sequent calculus, where the distinction between a principal succedent formula and the remaining succedent context can be examined directly.

Having demonstrated the preservation of a succedent formula under L∧ within singlesuccedent sequent calculus, we may now examine whether the same structural principle applies when an inference operates upon a formula occurring in a multisuccedent configuration. Consider the following application of the rule L¬:

$$\frac{\Gamma \Rightarrow \Delta, A}{\neg A, \Gamma \Rightarrow \Delta}\, L\neg$$

in which Γ and Δ represent the antecedent and succedent contexts, respectively. In this derivation, the application of L¬ removes A from the succedent and introduces its negation [¬A] into the antecedent, while preserving the remaining formulae represented by Γ and Δ.[7] Contrasting the preceding application of L∧, which operates exclusively upon antecedent formulae, L¬ involves a principal formula initially occurring in the succedent. However, the inferential transformation concerns only A, whereas the formulae represented by Δ retain their positions within the succedent of the resulting sequent. This distinction demonstrates that the preservation of conclusions is not confined to operations whose principal formulae occur in the antecedent. Rather, the multisuccedent configuration permits a logical operation to affect one succedent formula without requiring the remaining succedent formulae to undergo corresponding transformations. In particular, the transition from Γ ⇒ Δ,A ⊢ ¬A,Γ ⇒ Δ does not require any member of Δ to be independently derived again, since its continued occurrence is already included into the formulation of L¬. The significance of this example therefore lies not purely in the logical treatment of negation but in the structural relationship between the principal formula and the surrounding inferential context. While A no longer occurs in the succedent of the conclusion, the formulae represented by Δ remain available throughout the inference. This provides a further illustration of the proof-theoretic distinction between the transformation of a designated formula and the preservation of formulae unaffected by that transformation. Accordingly, the preceding applications of L∧ and L¬ demonstrate structural preservation under two different inferential configurations, preparing the ground for examining whether the same principle is maintained when an inference involves multiple premises.

The preservation of inferential context can also be demonstrated through the rule L∨:

$$\frac{\mathbf{A,\Gamma \Rightarrow \Delta \qquad B,\Gamma \Rightarrow \Delta}}{\mathbf{A \lor B,\Gamma \Rightarrow \Delta}}\ \mathbf{L\lor}$$

within multisuccedent sequent calculus, where a logical operation involving two premises retains their common succedent context. The rule L∨ combines the antecedent alternatives A and B into the disjunctive principal formula A ∨ B, while preserving the common antecedent context Γ and the succedent context Δ. Unlike the preceding applications of L∧ and L¬, this inference involves two premises, each containing a distinct antecedent formula upon which the disjunction rule operates. Yet, the transition to the conclusion does not alter the formulae represented by Δ, whose occurrence is maintained independently of the logical transformation performed upon A and B. The example therefore establishes that structural preservation is not restricted to single-premise inferences but also characterises rules whose applications depend upon multiple premises. Considered together, the three derivations involving L∧, L¬ and L∨ illustrate different configurations in which logical operations transform their designated formulae while preserving the remaining inferential context. The first demonstrates the preservation of a singlesuccedent conclusion during an antecedent transformation; the second shows that a multisuccedent rule may transform one succedent formula while preserving the others; and the third establishes that the common succedent context remains unchanged across a two-premise inference. These observations are significant for the proof-theoretic interpretation of reiteration because they identify structural preservation as a feature of the logical rules themselves rather than an additional operation introduced to maintain the availability of formulae. In natural deduction, the subsequent restatement of an available formula may require an explicit application of R, whereas the corresponding preservation in sequent calculus is represented through the contextual components of the inference. The examples thus provide the formal basis for investigating the two functions attributed to reiteration, namely the reuse of previously employed formulae and the restoration of previously derived conclusions. To examine these functions more precisely, we shall now introduce the schematic representation of R before comparing corresponding derivations in natural deduction and sequent calculus.

---

[7] As is our convention, formulae enclosed in [ ] in both natural deduction proofs and the body of this work are regarded as assumed formulae.

After examining the preservation of inferential contexts through the applications of L∧, L¬ and L∨, we may now consider how this structural behaviour provides a proof-theoretic foundation for the principle of reiteration in natural deduction. The principle, denoted by R, may be represented schematically as follows:

$$\dfrac{\mathbf{A}}{\mathbf{A}}\ \mathbf{R}$$

More generally, where Π denotes an arbitrary derivation of A:

$$\dfrac{\begin{array}{c}\mathbf{\Pi}\\ \mathbf{A}\end{array}}{\mathbf{A}}\ \mathbf{R}$$

These representations express the restatement of a formula already derived within a proof, rather than the derivation of a logically distinct conclusion. Unlike the introduction and elimination rules governing logical constants, R does not operate upon the internal logical structure of A or transform it into another formula. Its application instead makes explicit the continued availability of A at a subsequent stage of the derivation, provided that the relevant inferential dependencies remain in force.[8] This qualification is important because reiteration cannot legitimately restore a formula outside the scope of assumptions upon which its derivation depends. Within those dependencies, however, the previous employment of A in an inferential operation does not exhaust its availability or prevent its subsequent employment. Therefore, reiteration may serve two related purposes namely, permitting the reuse of formulae that have already participated in earlier inferences and restating previously derived conclusions after intervening operations. Although neither application contributes additional logical content, the explicit occurrence of R may be necessary to represent these relationships within a particular natural deduction derivation. The preceding sequent calculus examples suggest a structural explanation of this practice, since formulae outside the principal operation of a logical rule may be preserved through its contextual formulation without requiring independent rederivation. Nevertheless, the schematic representation of R alone does not establish the correspondence between the two proof environments. Such a correspondence must be examined through particular derivations in which the same inferential dependencies are represented differently. We shall therefore consider an example involving the rules →E and ∧I in natural deduction, together with the corresponding applications of the principles L→ and L∧ in sequent calculus.

[8] The continued availability of a formula must be distinguished from its unrestricted employment throughout a derivation. In particular, a formula derived under an assumption remains dependent upon that assumption and cannot be reiterated outside its legitimate scope following discharge. Therefore, reiteration preserves existing inferential dependencies rather than extending the domain within which a formula may be employed.

To examine the proposed correspondence, consider the following natural deduction derivation:

**A → B          [A]**
**____________________→E**
**B**
**____________________R**
**A → B          [A]**
**____________________∧I**
**(A → B) ∧ A**
**____________________R**
**B**

in which A → B and [A] are initially employed to derive B through →E. The same formulae are subsequently employed in ∧I, while B is finally restated through R. The corresponding sequent calculus derivation is represented as follows:

**A ⇒ A          B ⇒ B**
**_______________________L→**
**(A → B),A ⇒ B**
**_______________________L∧**
**(A → B) ∧ A ⇒ B**

The natural deduction example illustrates two distinct applications of reiteration. First, A → B and [A], having participated in the derivation of B through →E, remain available for subsequent employment in ∧I, which derives (A → B) ∧ A. Their earlier inferential participation therefore does not prevent their reuse, nor does their subsequent employment require their independent rederivation. Second, B, which was derived before the conjunction was introduced, is restated as the final conclusion without undergoing another logical transformation. These applications distinguish the reuse of previously employed formulae from the restoration of an earlier conclusion. The sequent calculus derivation represents the relevant structural relationships differently. Starting from the reflexive sequents A ⇒ A and B ⇒ B, the application of L→ derives (A → B),A ⇒ B, after which L∧ combines the antecedent formulae into (A → B) ∧ A while preserving B in the succedent. The continued occurrence of B therefore illustrates how a conclusion can remain structurally available despite an intervening logical operation upon the antecedent. Moreover, the participation of A → B and [A] in successive inferential configurations provides a corresponding representation of their availability for further logical operations. The comparison does not identify →E with L→ or ∧I with L∧ as identical rules; rather, it concerns the preservation of inferential dependencies across differently organised proof atmospheres.[9] In this respect, the derivations provide a concrete basis for examining the two functions of reiteration without attributing independent inferential content to R. But this correspondence raises questions concerning both the legitimacy of reusing previously employed formulae and the relationship between the different conjunction rules, which require further examination.

However, the proposed correspondence between natural deduction and sequent calculus raises two objections concerning the proof-theoretic legitimacy of reiteration and the relationship between their respective logical rules. The first objection concerns whether formulae that have already participated in an inferential operation may legitimately be employed again without being independently derived. In the preceding natural deduction derivation, A → B and [A] are initially employed through →E to derive B, after which they are reiterated for the subsequent application of ∧I. One might object that their earlier inferential employment has already fulfilled their function and that their reuse therefore requires an additional justification. This objection, nonetheless, presupposes that participation in an inference exhausts the subsequent availability of its premises, a presupposition that is not supported by the structural organisation of sequent calculus. In the corresponding derivation, the application of L→ to the reflexive sequents A ⇒ A and B ⇒ B produces (A → B),A ⇒ B, within which A → B and [A] remain available as antecedent formulae. Their participation in the preceding inference does not prevent the subsequent application of L∧, which derives (A → B) ∧ A ⇒ B while preserving the succedent B. The antecedent formulae are therefore available for further logical operations without requiring their independent rederivation. This demonstrates that inferential employment does not entail the consumption or elimination of the formulae involved, provided that their relevant dependencies remain available. As a result, the application of R in natural deduction may be understood as explicitly representing a structural continuity that sequent calculus includes into its inferential configuration. The objection thus fails to distinguish between the performance of a logical operation upon available formulae and the continued availability of those formulae for subsequent operations. While the former concerns the derivation of a conclusion according to a particular logical rule, the latter concerns the preservation of inferential resources within the derivation. This distinction is essential to the present account because it explains why reiteration permits the reuse of previously employed formulae without introducing an additional logical principle or extending the inferential content of the proof.

---

[9] The correspondence proposed here concerns the structural preservation of inferential dependencies rather than a rule-by-rule translation between natural deduction and sequent calculus. As a result, the comparison does not claim that the respective introduction, elimination and left rules perform identical logical operations. Its purpose is to demonstrate how the continued availability of formulae may be represented explicitly through reiteration in natural deduction and implicitly through preserved contexts in sequent calculus.

The second objection concerns the apparent discrepancy between the conjunction rules employed in the two derivations. In natural deduction, the formulae $A \rightarrow B$ and [A] are combined through ∧I to derive $(A \rightarrow B) \wedge A$, whereas the corresponding sequent calculus derivation employs L∧ rather than R∧. It might therefore be argued that the proposed correspondence is defective because the two rules occupy different inferential positions and perform distinct formal operations. This objection identifies a genuine difference between the proof environments, but it does not invalidate the structural relationship under examination. In natural deduction, ∧I derives a conjunctive conclusion from two available formulae, whereas L∧ in sequent calculus introduces a conjunction into the antecedent of the conclusion sequent from its constituent antecedent formulae in the premise. Consequently, the rules should not be identified as formally equivalent operations simply because they involve the same conjunctive formula. Their relevance to the present enquiry instead concerns the representation of inferential dependencies and the preservation of conclusions during subsequent logical operations. In the natural deduction derivation, the formula B, previously derived through →E, remains available after the intervening application of ∧I and is subsequently restated through R. In the sequent calculus derivation, B remains in the succedent throughout the application of L∧, notwithstanding the transformation of the antecedent formulae into $(A \rightarrow B) \wedge A$. The correspondence therefore does not require the conjunction rules to possess identical inferential functions, but only that the respective derivations exhibit the structural continuity necessary to explain the continued availability of B. Furthermore the availability of $A \rightarrow B$ and [A] for the subsequent conjunction operation illustrates how the two proof environments represent the reuse of inferential resources differently. Natural deduction makes this availability explicit through reiteration, while sequent calculus includes it into the antecedent structure of the derivation. Thus, the distinction between ∧I and L∧ does not undermine the proposed proof-theoretic explanation of reiteration, provided that the correspondence is understood structurally rather than as an identity between logical rules.

## 3. Concluding Remarks

This manuscript has examined the proof-theoretic foundations of reiteration in natural deduction by asking whether the continued availability of formulae requires an independent logical principle or can instead be explained through the structural organisation of sequent calculus. Our analysis supports the latter interpretation. The distinction between principal formulae and their surrounding inferential contexts shows that logical rules can transform designated formulae without eliminating other formulae upon which a derivation depends. For example, in singlesuccedent sequent calculus, the application of L∧ illustrates how an antecedent conjunction may be introduced while the succedent remains unchanged.[10] Equally, the multisuccedent applications of L¬ and L∨ extend this observation by demonstrating, respectively, the preservation of unaffected succedent formulae when one formula undergoes a negation operation and the maintenance of a common context across two premises.[11] These examples establish the structural setting within which reiteration can be understood without attributing to it a separate contribution to logical consequence. In natural deduction, the schematic application of R restates an available formula rather than deriving a distinct proposition. Its legitimacy consequently depends upon the continued validity of the inferential dependencies supporting that formula, including the scope of any assumptions involved. Reiteration cannot recover a conclusion after its supporting assumptions have ceased to be available. Accordingly, the principle expresses continuity within an existing derivation, not permission to disregard its dependency structure. The resulting account identifies structural preservation as the central explanatory feature linking reiteration to the contextual formulation of sequent calculus rules. This interpretation distinguishes derivational availability from logical transformation in general.

The comparison of particular derivations clarifies two functions of reiteration that must be distinguished. First, premises previously employed in an inference remain available for subsequent operations without requiring independent rederivation. Second, a conclusion derived before intervening operations may be restated when its supporting dependencies remain operative. The natural deduction example employing →E, ∧I and R illustrates both functions, while the corresponding sequent calculus derivation employing L→ and L∧ represents the preservation of antecedent resources and the succedent conclusion within its inferential structure. This correspondence is structural rather than an identification of the respective logical rules. It also answers two objections considered in the document. Earlier inferential employment does not, by itself, exhaust the availability of a formula; similarly, the difference between ∧I in natural deduction and L∧ in sequent calculus does not defeat a comparison concerned with preserved dependencies rather than identical operations. The discussion of double negation further illustrates why reiteration must be distinguished from reductio because restating an independently derived conclusion does not itself perform the logical operation associated with that derivation. Future research could investigate the precise conditions governing reiteration across subordinate proofs and assumption discharge, examine whether the proposed explanation admits a general formal correspondence between natural deduction and sequent calculus, and test its limits in linear, relevant and other substructural systems where weakening or contraction may be restricted. Such investigations would clarify both the scope and the limitations of reiteration as a structural principle. They may also illuminate questions concerning proof identity and normalisation.

---

[10] Of course, we acknowledge that this rule also holds in multisuccedent sequent calculus and that the present case serves merely as an example.

[11] As explained in the preceding footnote, these rules have the same effect in a singlesuccedent sequent calculus system and are just employed here as two examples to clarify our point.

## References

Alrubyli, Y. 2021. 'Natural Deduction Calculus for First-Order Logic', *arXiv:2108.06015v1*. https://doi.org/10.48550/arXiv.2108.06015

Fitch, F. B. 1952. *Symbolic Logic: An Introduction.* New York: Ronald Press Company.

Irani, K. 2025a. 'The Proof-Theoretic Origin of Double Negation Introduction & Elimination', (Under Review) *arXiv: 2509.17623V1*. https://doi.org/10.48550/arXiv.2509.17623

Irani, K. 2025b. 'Proof Analysis of A Foundational Classical Singlesuccedent Sequent Calculus', (Under Review) *arXiv:2507.04449v1*. https://doi.org/10.48550/arXiv.2507.04449

Irani, K. 2026a. 'The Nonlocal Nature of Natural Deduction & the Introduction of Assumptions in Derivations', (Under Review) *arXiv:2610.17076v1*. https://doi.org/10.48550/arXiv.2610.17076

Irani, K. 2026b. 'The Proof-Theoretic Validity of the Principles of Right Negation & Negation Introduction: A Generalised Theory of Reductio ad Absurdum', (Under Review) *Zenodo (2.0).* https://doi.org/10.5281/zenodo.23345562

Jang, J. Roshal, S. Pfenning, F. & Pientka, B. 2024. 'Adjoint Natural Deduction', In: *Proceedings of the 9th International Conference on Formal Structures for Computation and Deduction (FSCD 2024).* Leibniz International Proceedings in Informatics (LIPIcs). https://doi.org/10.4230/LIPIcs.FSCD.2024.15?utm_source=chatgpt.com)

Negri, S. & von Plato, J. 2001. *Structural Proof Theory*. Cambridge: Cambridge University Press. https://doi.org/10.1017/cbo9780511527340

Negri, S. & Von Plato, J. 2011. *Proof Analysis: A Contribution to Hilbert's Last Problem.* Cambridge: Cambridge University Press. https://doi.org/10.1017/CBO9781139003513

Pelletier, F. J. & Hazen, A. 2026. 'Natural Deduction Systems in Logic', in Zalta, E. N. & Nodelman, U. (eds.) *The Stanford Encyclopedia of Philosophy.* Accessed on 25/10/2026. https://plato.stanford.edu/entries/natural-deduction/

Prawitz, D. 1965. *Natural Deduction: A Proof-Theoretical Study.* Stockholm: Almqvist & Wiksell.

Prawitz, D. 1971. 'Ideas and Results in Proof Theory', *Journal of Symbolic Logic.* Vol. 40. No. 2. pp. 232-305. https://doi.org/10.1016/s0049-237x(08)70849-8

Restall, G. 2023. 'Structural Rules in Natural Deduction with Alternatives', *Bulletin of the Section of Logic.* Vol. 52. No. 2. pp. 109-143. https://doi.org/10.18778/0138-0680.2023.6

Smiley, T. & Shoesmith, D. J. 1978. *Multiple-Conclusion Logic.* Cambridge: Cambridge University Press. https://doi.org/10.1017/cbo9780511565687

Szabo, M. E. (ed.) 1969. *The Collected Papers of Gerhard Gentzen.* Amsterdam: North-Holland Publishing Company.

Troelstra, A. S. & Schwichtenberg, H. 2000. *Basic Proof Theory.* Cambridge: Cambridge University Press. https://doi.org/10.1017/cbo9781139168717